\documentclass[aip,graphicx, reprint]{revtex4-1}

\usepackage{graphicx}
\usepackage{dcolumn}
\usepackage{bm}
\usepackage[version=3]{mhchem} 
\usepackage[utf8]{inputenc}
\usepackage[T1]{fontenc}
\usepackage{mathptmx}
\usepackage{etoolbox}
\usepackage{siunitx}
\usepackage{amsmath}
\usepackage{lipsum}
\usepackage{placeins}

\usepackage[dvipsnames,svgnames,x11names]{xcolor}
\usepackage[markup=added, authormarkup=none]{changes}  

\definechangesauthor[name={}, color=NavyBlue]{BK}
\definechangesauthor[name={}, color=BrickRed]{YW}

\draft 

\begin{document}


\title{62.6 GHz ScAlN Solidly Mounted Acoustic Resonators } 



\author{Yinan Wang}
\thanks{These authors contributed equally to this work.}
\affiliation{University of Texas at Austin, Austin, TX 78758, USA}

\author{Byeongjin Kim}
\thanks{These authors contributed equally to this work.}
\affiliation{University of Texas at Austin, Austin, TX 78758, USA}

\author{Nishanth Ravi}
\affiliation{University of California at Los Angeles, Los Angeles, CA 90095, USA}

\author{Kapil Saha}
\affiliation{Northeastern University, Boston, MA 02115, USA}

\author{Supratik Dasgupta}
\affiliation{Bühler Leybold Optics, Cary, NC 27513, USA}

\author{Vakhtang Chulukhadze}
\affiliation{University of Texas at Austin, Austin, TX 78758, USA}

\author{Eugene Kwon}
\affiliation{University of California at Los Angeles, Los Angeles, CA 90095, USA}

\author{Lezli Matto}
\affiliation{University of California at Los Angeles, Los Angeles, CA 90095, USA}

\author{Pietro Simeoni}
\affiliation{Northeastern University, Boston, MA 02115, USA}

\author{Omar Barrera}
\affiliation{University of Texas at Austin, Austin, TX 78758, USA}

\author{Ian Anderson}
\affiliation{University of Texas at Austin, Austin, TX 78758, USA}

\author{Tzu-Hsuan Hsu}
\affiliation{University of Texas at Austin, Austin, TX 78758, USA}

\author{Jue Hou}
\affiliation{Bühler Leybold Optics, Cary, NC 27513, USA}

\author{Matteo Rinaldi}
\affiliation{Northeastern University, Boston, MA 02115, USA}

\author{Mark S. Goorsky}
\affiliation{University of California at Los Angeles, Los Angeles, CA 90095, USA}

\author{Ruochen Lu}
\email{ruochen@utexas.edu}
\affiliation{University of Texas at Austin, Austin, TX 78758, USA}


\date{\today}

\begin{abstract}
We demonstrate a record-high \SI{62.6}{GHz} solidly mounted acoustic resonator (SMR) incorporating a \SI{67.6}{nm} scandium aluminum nitride (\ce{Sc_{0.3}Al_{0.7}N}) piezoelectric layer on a \SI{40}{nm} buried platinum (\ce{Pt}) bottom electrode, positioned above an acoustic Bragg reflector composed of alternating \ce{SiO2} (\SI{28.2}{nm}) and \ce{Ta2O5} (\SI{24.3}{nm}) layers in 8.5 pairs. The Bragg reflector and piezoelectric stack above are designed to confine a third-order thickness-extensional (TE) bulk acoustic wave (BAW) mode, while efficiently transducing with thickness-field excitation. The fabricated SMR exhibits an extracted piezoelectric coupling coefficient ($k^2$) of 0.8\% and a maximum Bode quality factor ($Q$) of 51 at \SI{63}{GHz}, representing the highest operating frequency reported for an SMR to date. These results establish a pathway toward mmWave SMR devices for filters and resonators in next-generation RF front ends.
\end{abstract}

\maketitle 


Piezoelectric acoustic-wave devices play a crucial role in state-of-the-art sub-6 GHz RF front-end filters due to their compact footprint, low insertion loss, and high frequency selectivity compared to electromagnetic (EM) counterparts.\cite{ 991846, 4803350, 9318743} As wireless systems advance toward 5G/6G and millimeter-wave (mmWave, i.e., above 30 GHz) bands, scaling acoustic filters to higher frequencies is desired,\cite{8052573, 9265094} but also presents substantial challenges. The challenges stem from the moderate performance of incumbent piezoelectric resonators at mmWave.\cite{10187967, 9552228, 5285627} Since filters are built from electrically coupled resonators, the performance of the resonator directly bounds the bandwidth, loss, and linearity of the filter. \cite{11242218, 11005482, 10989534} One key acoustic platform, surface acoustic wave (SAW) devices, sets resonant frequency through the interdigitated transducer (IDT) pitch,\cite{9122391, 8536421, 10617811} but pushing IDT dimensions into the mmWave regime increases fabrication complexity and exacerbates loss mechanisms.\cite{10.1063/1.3475987, 10052227} Another technology, bulk acoustic wave (BAW) devices, scales resonant frequency inversely with film thickness.\cite{991846, 8614564} However, forcing the fundamental mode into mmWave often demands sub-100-nm films, which tend to get degraded material quality and elevated loss.\cite{LU2025100565, Lu_2021} Exploiting higher-order thickness-extension (TE) modes offers a route to achieving performance in the mmWave range without requiring extreme thinning.\cite{9994579} For instance, thin-film bulk acoustic resonators (FBARs) in sputtered \ce{ScAlN} with bottom metal electrodes have reached ~\SI{60}{GHz} using third-order TE modes,\cite{10439299} but suspended FBARs face limited power handling and structure integrity at extreme miniaturization.\cite{10153792} 

A promising alternative is solidly mounted resonators (SMRs), which enhance power and structural robustness while still confining acoustic energy with an acoustic Bragg reflector,\cite{495711, 4409835} rather than air in the suspended BAW counterparts. The materials and thicknesses in the Bragg reflector can be adjusted to target specific frequencies. These advantages make SMRs an attractive platform for high-frequency acoustic resonators. However, most reported SMRs so far operate at frequencies below a few tens of gigahertz (Table~\ref{tab:soa}), and maintaining both high quality factor ($Q$) and coupling ($k^2$) at higher frequencies becomes difficult as the acoustic wavelength shrinks. A few demonstrations above \SI{10}{GHz} exhibit limited $k^2$ (e.g., $\le$3\%) due to lateral-field excitation and incomplete mode confinement.\cite{10718726} Some recent works show all-metal Bragg reflectors at 50 GHz.\cite{10307494,11083120} However, these devices rely on a rather complicated fabrication flow, and major challenges in mitigating the feedthrough from metal reflectors. Overcoming these limits requires co-optimizing the Bragg stack and electrode configuration to confine higher-order BAW modes while leveraging thickness-field excitation.

\begin{table}[t]
\caption{State-of-the-art SMR resonators.}
\label{tab:soa}
\centering
\begin{ruledtabular}
\begin{tabular}{c c c c c c}
Reference & Mirror Stack & $f$ (GHz) & $Q$ & $k^{2}$ (\%) & $f\!\cdot\!Q$ ($\times 10^{12}$) \\
\hline
\textit{Kadota et al.\cite{10306999}}    & \ce{SiO2\text{/}Ta}       & 9.5  & 400   & 2.0  & 3.8  \\
\textit{Lv et al.\cite{doi:10.1021/acsomega.2c01749}}        & \ce{SiO2\text{/}Ta2O5}    & 3.5  & 225   & 17.9 & 0.79 \\
\textit{Tag et al.\cite{9958625}}       & \ce{SiO2\text{/}W}        & 7.5  & 2500  & N/A  & 18.8 \\
\textit{Kimura et al.\cite{8617704}}    & N/A                & 4.9  & 565   & 24.0 & 2.8  \\
\textit{Schaffer et al.\cite{10307494}}  & \ce{Al\text{/}W}          & 55.7 & 95    & 2.2  & 5.3  \\
\textit{Baek et al.\cite{11083120}}    & \ce{Al\text{/}W}  & 51.3  & 108   & 6.1  & 5.5  \\
\textit{Bousquet et al.\cite{10308137}}  & \ce{SiO2\text{/}AlN}      & 4.8  & 560   & 12.7 & 2.7  \\
\textit{Barrera et al.\cite{10718726}}   & \ce{SiO2\text{/}Ta2O5}
    & 18.6 & 210   & 2.0  & 3.9  \\
\textit{Anderson et al.\cite{anderson2025solidly}}   & \ce{SiO2\text{/}Nb2O5}
    & 18.6 & 205   & 2.0  & 3.8  \\
\textit{Anderson et al.\cite{anderson2025solidly}}   & \ce{SiO2\text{/}Ta2O5}
    & 18.6 & 206   & 2.5  & 3.8  \\
\textbf{This work}        & \ce{SiO2\text{/}Ta2O5}    & 62.6 & 51 & 0.8  & 3.2 \\
\end{tabular}
\end{ruledtabular}
\end{table}

In this work, we demonstrate a \SI{63}{GHz} SMR based on \ce{Sc_{0.3}Al_{0.7}N} with a \SI{67.6}{nm} active layer and a buried \SI{40}{nm} \ce{Pt} bottom electrode above a \ce{SiO2}/\ce{Ta2O5} Bragg reflector (8.5 pairs, \SI{28.2}{nm}/\SI{24.3}{nm}). Finite-element analysis (FEA) guides the stack design to confine and efficiently excite the third-order TE mode. The fabricated devices achieve $k^2=\SI{0.8}{\%}$ and Bode $Q$ of 51 at \SI{63}{GHz}, representing, to our knowledge, the highest operating frequency reported for an SMR to date. These results outline a pathway to compact mmWave filters and resonators for next-generation RF front ends.

\begin{figure}[t]
    \centering
    \added{\includegraphics[width=\columnwidth]{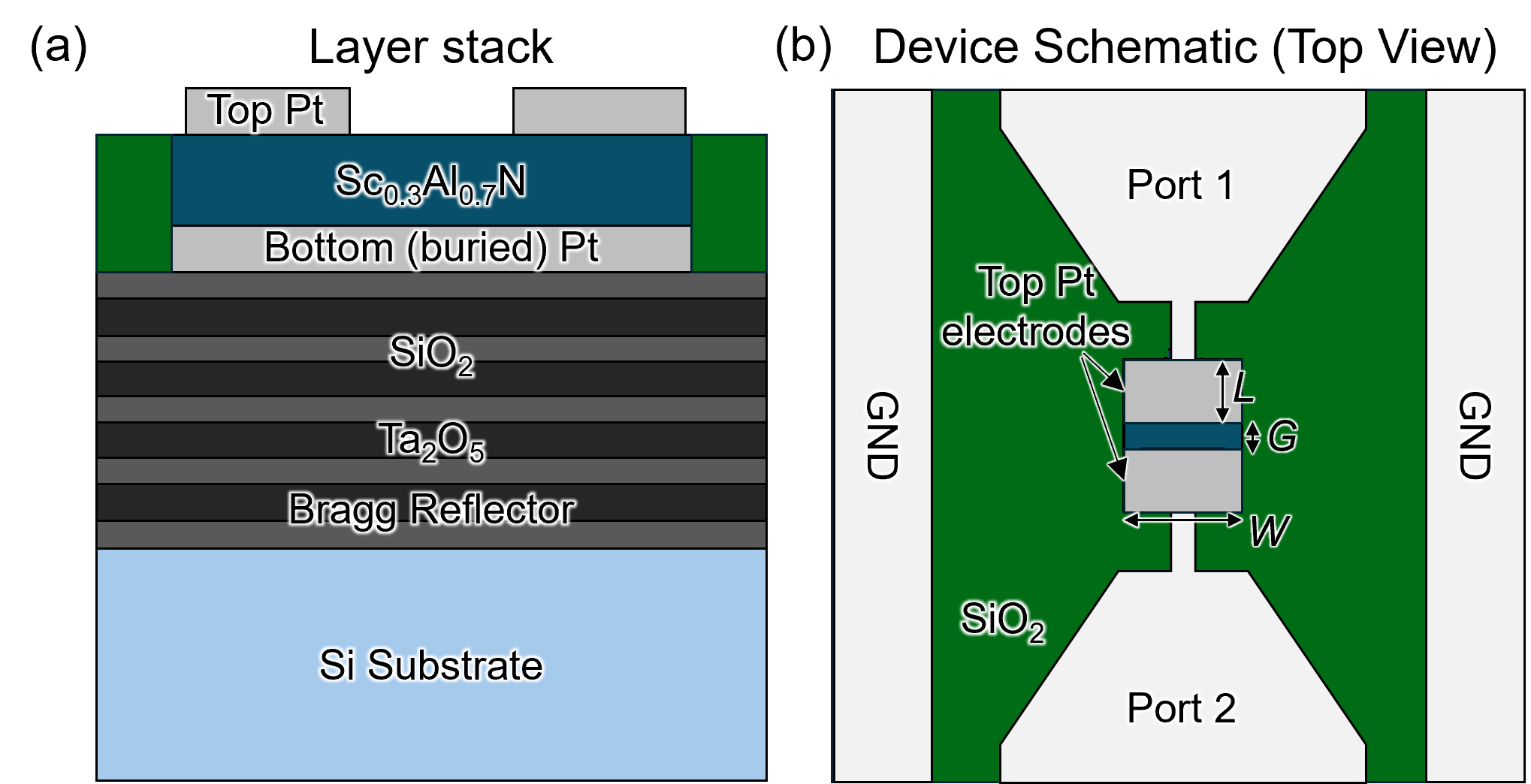}}
    \caption{SMR structure, device layout, and key dimensions. (a) Cross-sectional schematic of the Pt/ScAlN/Pt stack on a \ce{SiO2}/\ce{Ta2O5} Bragg reflector on a high-resistivity Si substrate. In the Bragg reflector, \ce{SiO2} is shown in light gray and \ce{Ta2O5} in dark gray; only a subset of the 8.5 pairs is drawn for clarity. (b) Top-view schematic of the device layout. Dimensions are listed in Table II. 
    }
    \label{fig:stack}
\end{figure}

\begin{table}[t]
\caption{Key device parameters.}
\label{tab:device_params}
\centering
\setlength{\tabcolsep}{3.5pt}%
\renewcommand{\arraystretch}{1.05}%
\begin{ruledtabular}
\begin{tabular}{c c c c c c}
Sym. & Parameter & Value & Sym. & Parameter & Value \\
\hline
$t_{\mathrm{Pt}}$     & Pt thk.         & \SI{40}{nm}            & $L$                & Elec. L     & \SI{8.5}{\micro\meter} \\
$t_{\mathrm{ScAlN}}$  & \ce{ScAlN} thk. & \SI{67.6}{nm}          & $W$                & Elec. W     & \SI{8.5}{\micro\meter} \\
$t_{\mathrm{SiO2}}$   & \ce{SiO2} thk.  & \SI{28.2}{nm}          & $G$                & Gap         & \SI{4.2}{\micro\meter} \\
$t_{\mathrm{Ta2O5}}$  & \ce{Ta2O5} thk. & \SI{24.3}{nm}          & $N_{\mathrm{refl}}$ & Refl. pairs & 8.5 \\
\end{tabular}
\end{ruledtabular}
\end{table}

Fig.~\ref{fig:stack} summarizes the SMR stack and device layout. From the substrate upward, the structure comprises an acoustic Bragg reflector and a piezoelectric resonator cavity [Fig.~\ref{fig:stack}(a)]. The reflector is formed by 8.5 pairs of alternating \ce{SiO2} (\SI{28.2}{nm}) and \ce{Ta2O5} (\SI{24.3}{nm}), terminated on \ce{SiO2}, on a high-resistivity Si (HR–Si) substrate. Although simulations suggest that fewer \ce{SiO2{/}Ta2O5} pairs can provide adequate isolation under nominal material parameters (Supplementary Fig. 1), we selected 8.5 pairs to increase robustness against thin‑film property variations and process tolerances while remaining compatible with the deposition and subsequent etching constraints. Above it, a buried \ce{Pt} bottom electrode (\SI{40}{nm}), a \ce{Sc_{0.3}Al_{0.7}N} active layer (\SI{67.6}{nm}), and a patterned \ce{Pt} top electrode define the FBAR region. In operation, the Bragg reflector transforms the Si substrate impedance to a low effective value, approximating a free boundary and confining bulk acoustic motion above the reflector.

The \ce{ScAlN} and \ce{Pt} thicknesses are chosen to support the third-order TE mode while preserving strong thickness-field coupling by placing approximately a half acoustic wavelength in the top and bottom \ce{Pt} electrodes and in the \ce{ScAlN} film. This alignment positions a stress antinode at the \ce{Pt}/\ce{ScAlN} interface, maximizing $e_{33}$-mediated coupling and minimizing acoustic leakage into the substrate. The selected thicknesses are consistent with prior suspended ScAlN-Pt BAW resonators at 50 GHz.\cite{10439299} A pair of top \ce{Pt} electrodes routes the mmWave signals [Fig.~\ref{fig:stack}(b)]. The buried floating bottom electrode establishes a strong thickness-directed electric field in \ce{ScAlN} and leveraging $e_{33}$. To suppress pad feedthrough capacitance, \ce{ScAlN} and \ce{Pt} outside the active area are removed and backfilled with \ce{SiO2} for isolation. 

Here, we select an acoustic Bragg stack composed entirely of all-dielectric materials, without using metals, as typically used in previously reported SMRs. It primarily reduces capacitive feedthrough, which is important for high-capacitance-density mmWave SMRs. In addition to reduced capacitive feedthrough, the SMR architecture is mechanically more robust than suspended FBARs and can potentially offer improved power handling due to the solid heat-sinking path into the substrate. While the all-dielectric \ce{SiO2{/}Ta2O5} stack is selected here to minimize RF feedthrough, alternative reflector designs employing higher-thermal-conductivity layers, including hybrid or all-metal Bragg reflectors \cite{10307494, 11083120}, may further improve heat spreading. These thermal considerations will be explored in future work. It also simplifies achieving uniform thickness with available deposition tools. Among dielectric materials, the reflector materials were chosen for their high acoustic impedance contrast, where \ce{SiO2} is the low-impedance layer ($Z_{\mathrm{SiO2}}=12.4~\mathrm{Mkg\,m^{-2}\,s^{-1}}$, $v_{\mathrm{SiO2}}=\SI{5640}{m/s}$) and \ce{Ta2O5} is the high-impedance layer ($Z_{\mathrm{Ta2O5}}=33.3~\mathrm{Mkg\,m^{-2}\,s^{-1}}$, $v_{\mathrm{Ta2O5}}=\SI{4860}{m/s}$).\cite{10718726} For a two-material Bragg stack, the fractional stopband bandwidth is approximated as \(\mathrm{FBW}=\frac{4}{\pi}\arcsin\!\left(\frac{Z_2-Z_1}{Z_2+Z_1}\right)\), where $Z_1$ and $Z_2$ are the longitudinal acoustic impedances of the low- and high-$Z$ layers. Substituting the values above gives $\mathrm{FBW}=60\%$, providing robust confinement even with process tolerances. The thicknesses of the Bragg layers are calculated based on quarter-wavelengths at 50 GHz, giving $t_{\mathrm{SiO2}}=\SI{28.2}{nm}$ 
and $t_{\mathrm{Ta2O5}}=\SI{24.3}{nm}$. Accordingly, the reflector uses \SI{28.2}{nm} \ce{SiO2} and \SI{24.3}{nm} \ce{Ta2O5} per pair to center the stopband on the third-order TE resonance of the \ce{Pt}/\ce{ScAlN}/\ce{Pt} cavity. 

\begin{figure}[t]
    \centering
    \includegraphics[width=0.95\columnwidth]{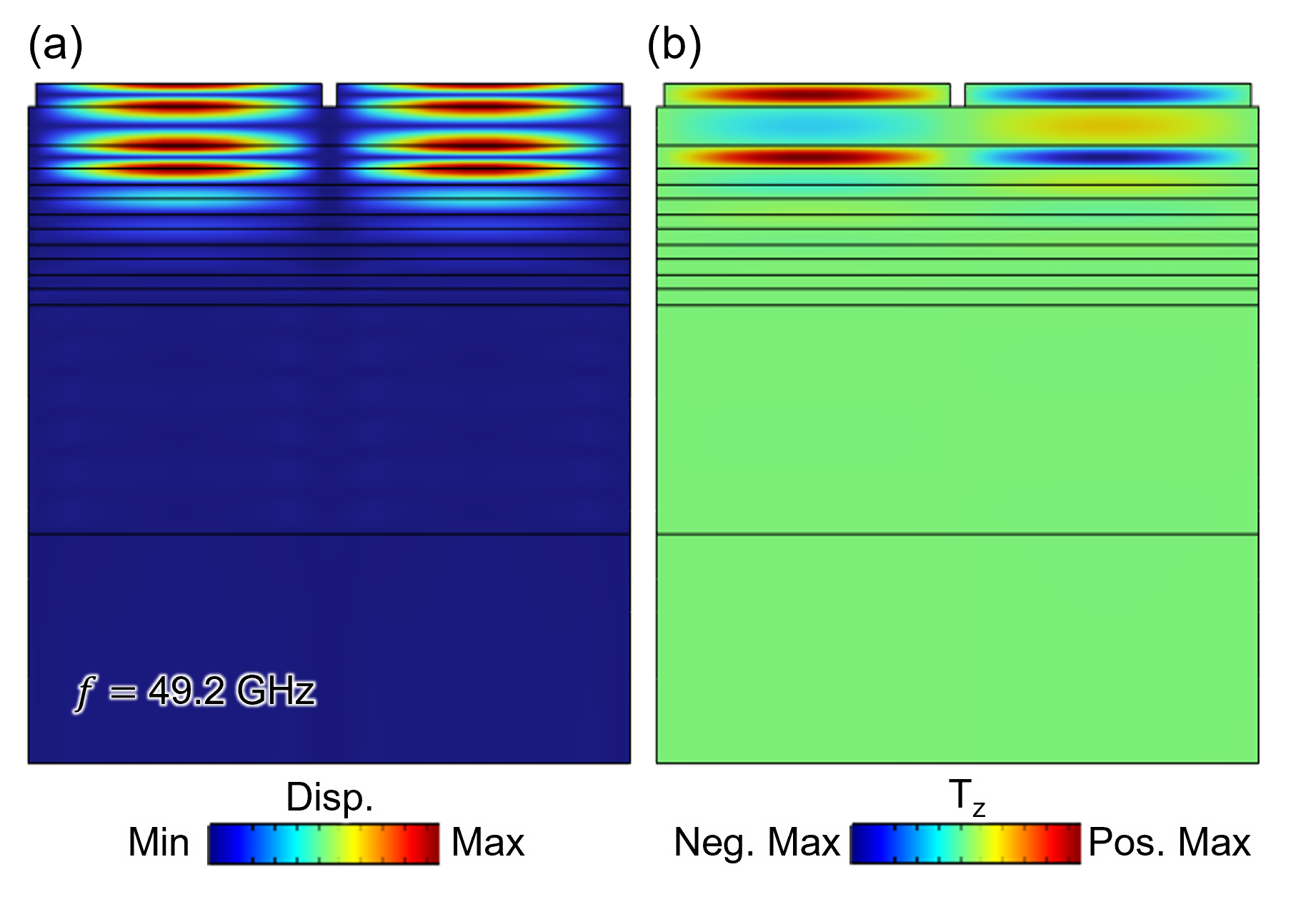}
    \caption{FEA mode shape at third-order TE resonance of 49.2 GHz. (a) Displacement confined to the \ce{Pt}/\ce{ScAlN}/\ce{Pt} cavity with exponential decay into the \ce{SiO2}/\ce{Ta2O5} reflector. (b) Axial stress $T_z$ with alternating sign and stress antinodes at the \ce{Pt}/\ce{ScAlN} interfaces.}
    \label{fig:mode}\label{fig:stress}
\end{figure}

To further confirm the design, the structure with the dimensions listed in Table~\ref{tab:device_params} was modeled in COMSOL three-dimensional (3D) eigenmode FEA. Lateral boundaries are set to periodic conditions to eliminate lateral dimension effects; the top surface is mechanically free; the substrate side uses a perfectly matched layer (PML) beneath the Bragg reflector to absorb any potential residual leakage into the HR-Si. Other than the PML, the structure is assumed to be lossless. The eigenmode FEA yields the targeted mode at 49.2 GHz (Fig.~\ref{fig:mode}), where the displacement is strongly confined to the \ce{ScAlN}/Pt stack and decays exponentially into the Bragg reflector [Fig.~\ref{fig:mode}(a)]. The higher-order TE shows multiple displacement antinodes across the \ce{Pt}/\ce{ScAlN}/\ce{Pt} cavity. The corresponding axial stress ($T_z$) distribution [Fig.~\ref{fig:stress}(b)] reinforces thickness-field excitation via $e_{33}$. The stress decay follows the same periodicity as the reflector pairs, validating the design.

\begin{figure}[t]
    \centering
    \includegraphics[width=0.95\columnwidth]{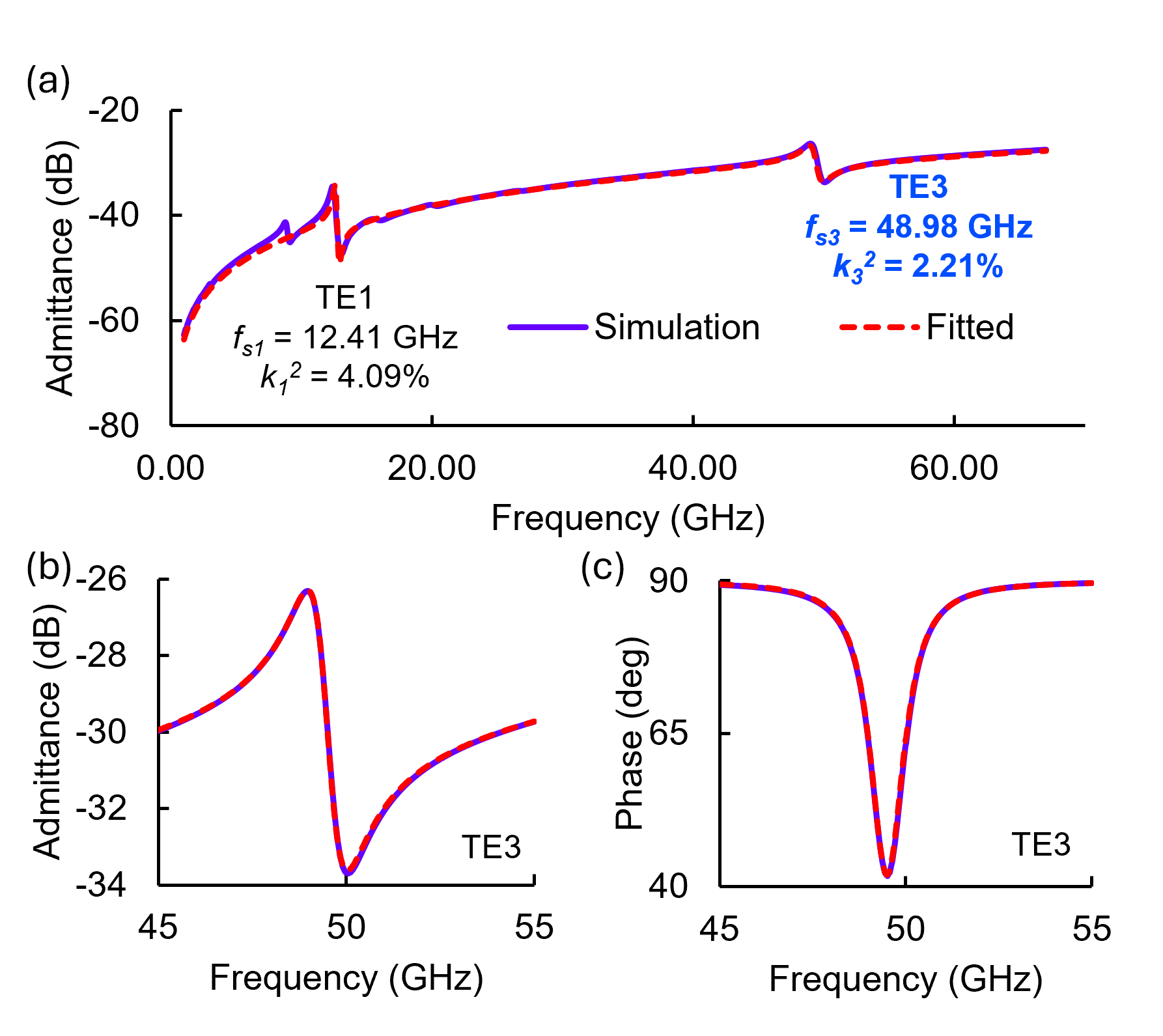}
    \caption{Simulated admittance of the SMR. (a) Simulated wideband admittance magnitude and key extracted parameters. (b-c) Zoomed-in admittance (b) magnitude and (c) phase around \SI{50}{GHz}.}
    \label{fig:admittance}
\end{figure}

The frequency-domain FEA is then performed for the same structure. A mechanical damping ($Q$ of 50) is applied in the model, based on prior measurements of ScAlN resonators at similar frequencies.\cite{10439299} The wideband admittance $Y(f)$ shows first and third order TE modes at 12.4 GHz and 49 GHz, respectively [Fig.~\ref{fig:admittance}(a)]. The zoomed-in admittance of the targeted third-order TE mode at 49 GHz is shown in Fig.~\ref{fig:admittance}(b) and (c). A multi-motional Butterworth–Van Dyke (BVD) fit \cite{8624370} is adapted, yielding an effective $k^{2}$ of 2.21\%. These results confirm that the design achieves the intended mode confinement and coupling for mmWave SMR operation.

\begin{figure}[t]
    \centering
    \added{\includegraphics[width=0.95\columnwidth]{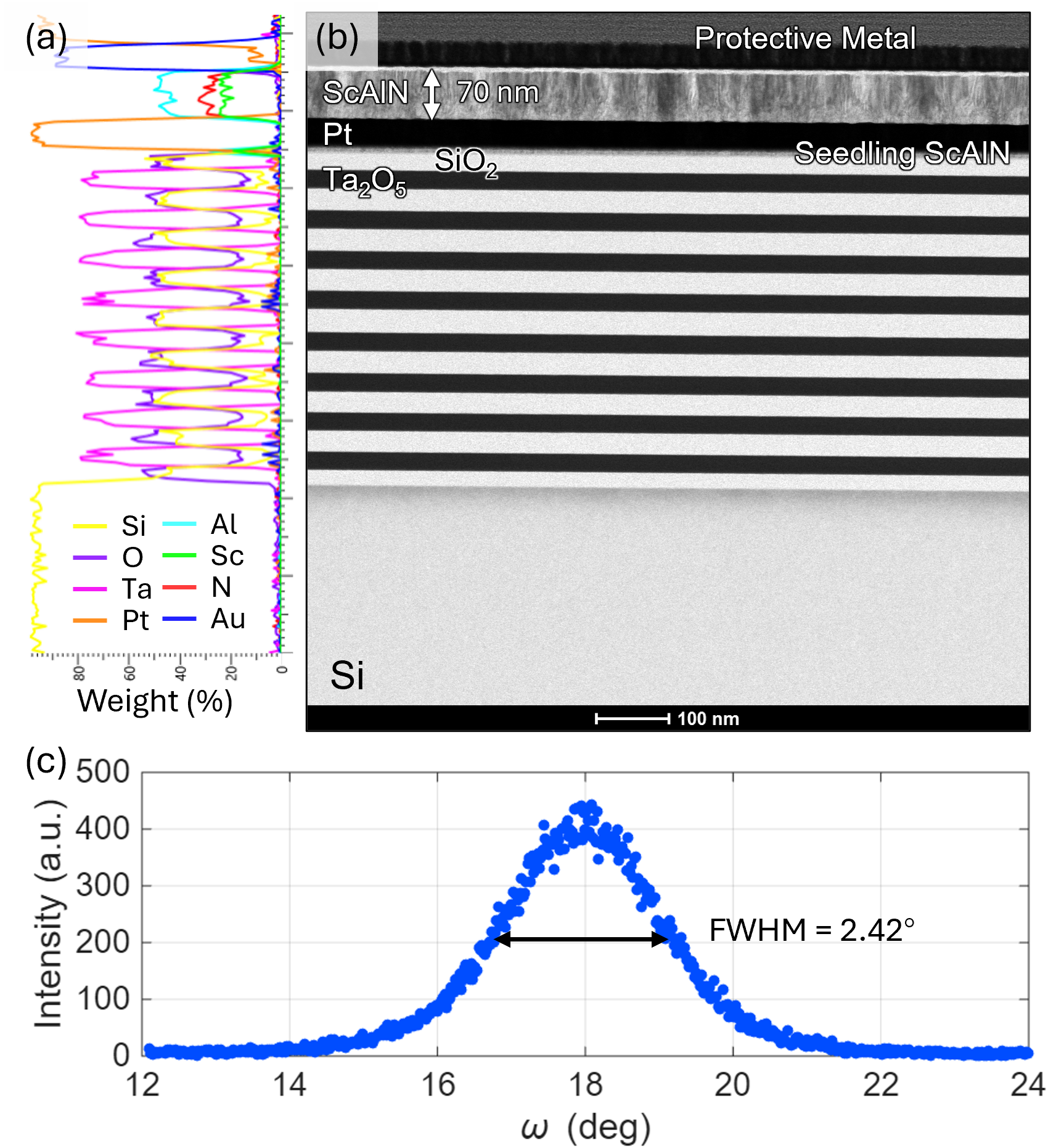}}
    \caption{Structural verification of the SMR stack. (a) EDS line-scan confirming the expected compositional periodicity of the \ce{SiO2{/}Ta2O5} Bragg reflector and the Pt/ScAlN/Pt structure. The top protective metal is used only for TEM sample preparation. (b) Cross-sectional TEM of the layer stack. (c) XRD showing strong c-axis orientation of the ScAlN film. Note that the protective metal layer in Fig. 4(b) is deposited only during TEM sample preparation and is not the patterned top Pt electrode in the measured devices.
}
    \label{fig:metrology}
\end{figure}

\begin{figure}[t]
    \centering
    \added{\includegraphics[width=0.95\columnwidth]{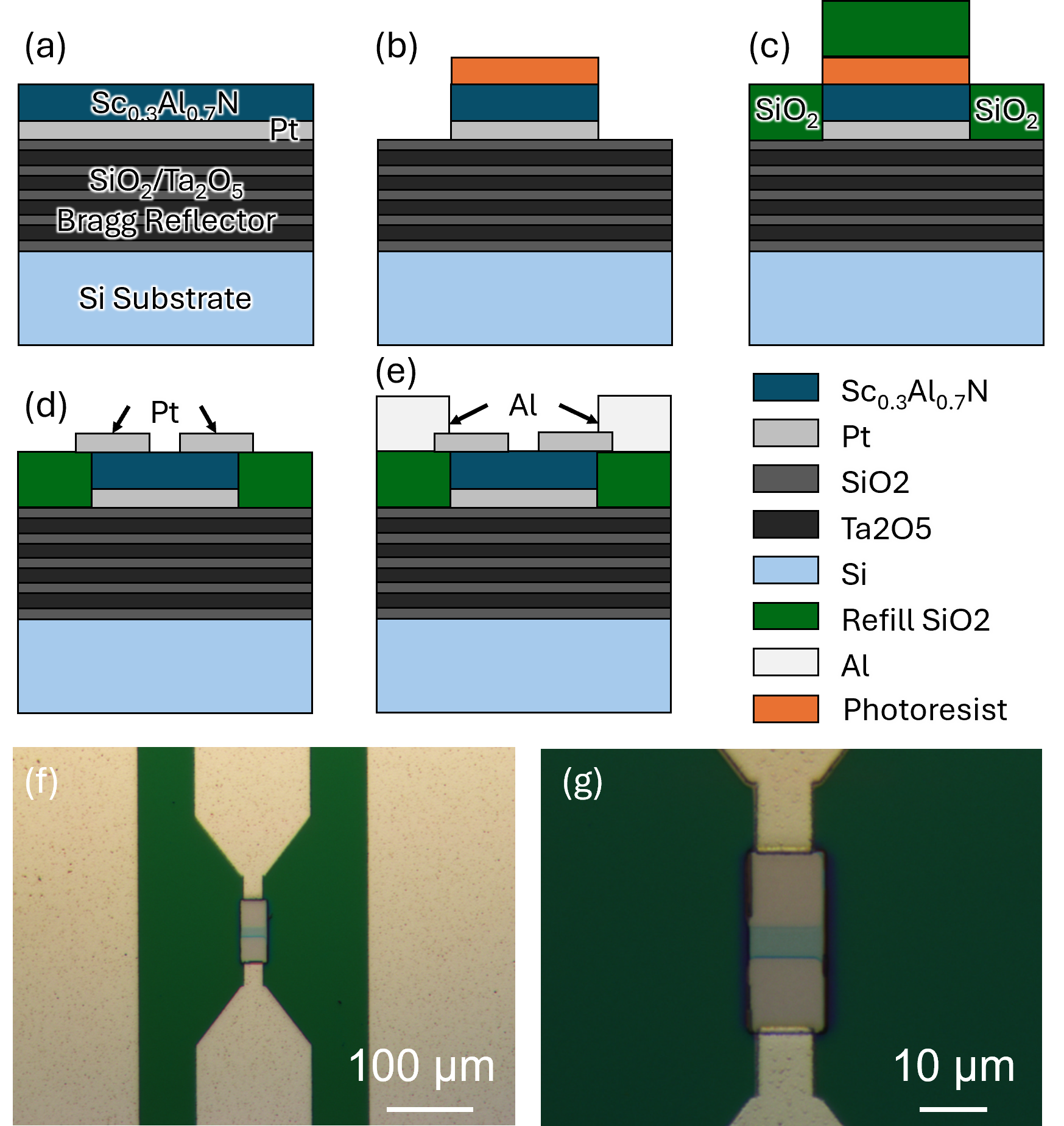}}
    \caption{Fabrication flow and device images. (a) As-deposited layer stack of \ce{Sc_{0.3}Al_{0.7}N} on buried Pt on \ce{SiO2{/}Ta2O5} Bragg reflector on HR-Si. Only a subset of the Bragg reflector pairs is illustrated (schematic not to scale). (b) Mesa definition and ion-mill etching of \ce{ScAlN/Pt}. (c) Low-temperature PECVD \ce{SiO2} backfill and lift-off. The PECVD \ce{SiO2} backfill/isolation oxide is shown in green to distinguish it from the Bragg-stack \ce{SiO2} layers. (d) Top Pt electrode patterning and deposition. (e) Thick electrode metal (\SI{300}{nm} Al) deposition. (f) Optical microscope image of a completed SMR device. (g) Zoomed-in optical image of the resonator region.}
    \label{fig:fabrication}
\end{figure}

The design is experimentally implemented as follows. Starting from an HR-Si substrate, the 8.5 pair \ce{SiO2}/\ce{Ta2O5} acoustic Bragg reflector was deposited using Helios 800 sputter coater (Bühler Leybold Optics) equipped with an OMS (Optical Monitoring System). A detailed description of the process can be found in a previous work that uses the same process.\cite{10718726} Following the reflector, a buried \ce{Pt} bottom electrode of \SI{40}{nm} was sputtered, and a \SI{67.6}{nm} \ce{Sc_{0.3}Al_{0.7}N} active layer was deposited in an Evatec Clusterline-200 magnetron system using a 12-inch \ce{Sc_{0.3}Al_{0.7}N} cast target. 

The quality and fidelity of the layer stack are verified using metrology methods, including cross-sectional transmission electron microscopy (TEM), energy-dispersive X-ray spectroscopy (EDS), and X-ray diffraction (XRD), and shown in Fig.~\ref{fig:metrology}. The EDS line scan [Fig.~\ref{fig:metrology}(a)] is shown side-by-side with the TEM image, confirming the as-expected composition of the Bragg reflectors, as well as the buried Pt and piezoelectric \ce{ScAlN} layers. The TEM image [Fig.~\ref{fig:metrology}(b)] shows distinct interfaces between each individual layer over all 8.5 periods of the Bragg reflector and a uniform \ce{ScAlN} layer bounded by the \ce{Pt} electrodes, consistent with the design. Lastly, the XRD scan [Fig.~\ref{fig:metrology}(c)] confirms the c-axis orientation of the \ce{ScAlN} layer with a full-width-half-maximum (FWHM) of 2.42$^{\circ}$.

The device fabrication flow chart can be found in [Fig.~\ref{fig:fabrication}(a)-(e)]. Device fabrication starts with a single-mask mesa etch and backfill sequence. The active region was defined by UV photolithography and patterned by ion milling to etch away \ce{ScAlN} and the buried \ce{Pt} outside of the region of interest. Without stripping the resist, a layer of \ce{SiO2} backfill was deposited by PECVD at \SI{100}{\celsius}; lift-off produced a self-aligned planarization and isolation around the resonator. The top \ce{Pt} electrode (targeting \SI{40}{nm} to match the buried layer) was patterned using UV photolithography, electron-beam evaporation, and lift-off. Finally, a \SI{300}{nm} \ce{Al} was deposited in the contact pad region to reduce probe resistance. The optical images of the fabricated device and zoomed-in active regions are shown in Fig.~\ref{fig:fabrication}(f) and (g), respectively.

\begin{figure}[t]
    \centering
    \added{\includegraphics[width=\columnwidth]{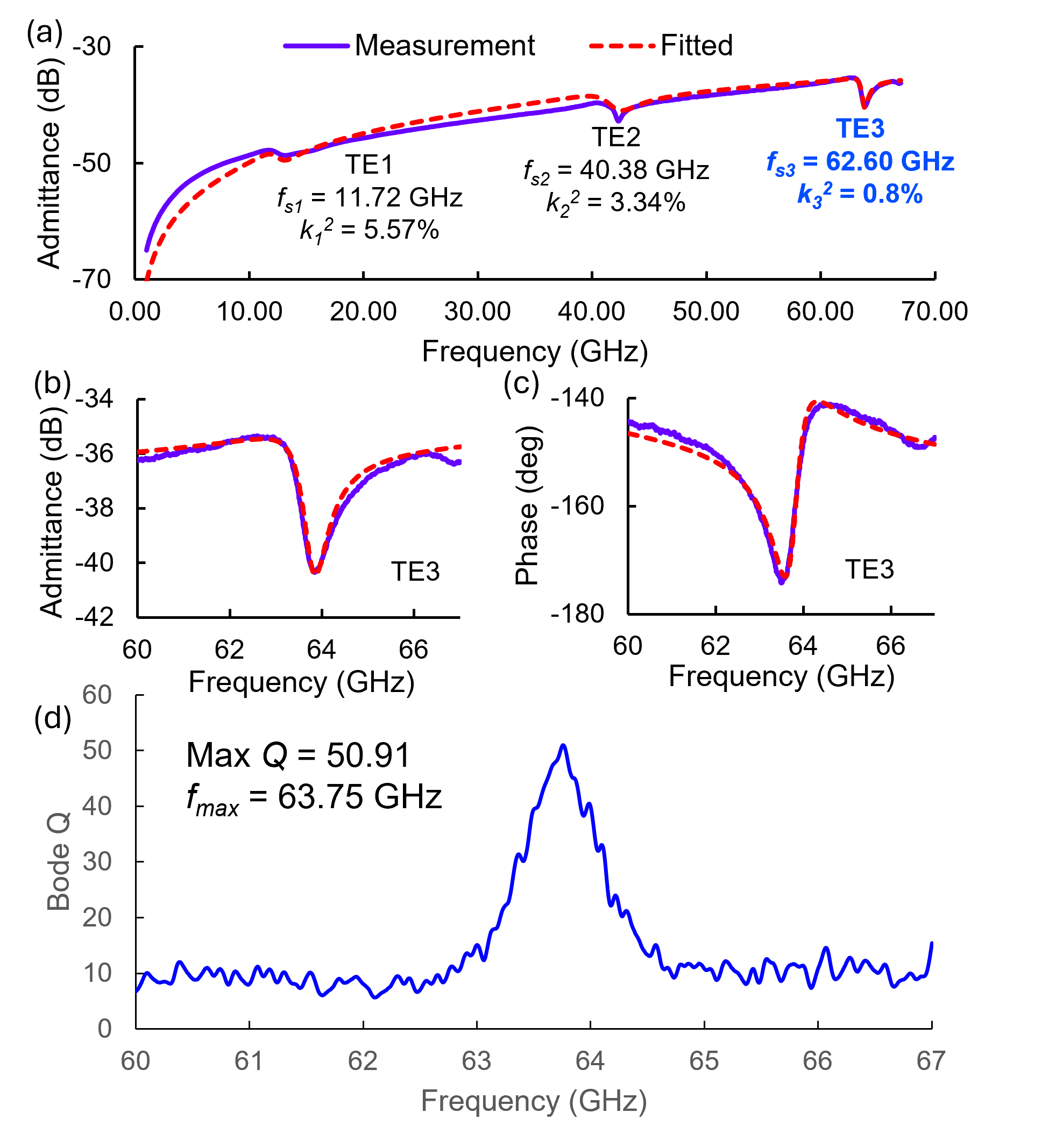}}
    \caption{Measured response of the SMR. (a) Measured wideband admittance magnitude and phase. (b) Zoomed-in admittance amplitude and (c) phase around the high-frequency TE3 mode $f_s$ and $f_p$. (d) Bode $Q$ peaking at $Q=50.91$ at \SI{63.75}{GHz}.}
    \label{fig:admittance_meas}
\end{figure}

The fabricated devices were characterized using GGB ground–signal–ground (GSG) probes connected to a 67 GHz Keysight vector network analyzer (VNA). The VNA is calibrated to the probe plane using a standard GGB CS-5 calibration substrate. Two-port $S$-parameters were measured from \SI{1} to \SI{67}{GHz} and converted to $Y$-parameters. Devices were measured as a series element; no pad de‑embedding was applied beyond CS‑5 calibration. Figure~\ref{fig:admittance_meas}(a) shows the measured wideband admittance. A clear series resonance appears at $\SI{62.6}{GHz}$ followed by an antiresonance at $\SI{63.8}{GHz}$. Figures~\ref{fig:admittance_meas}(b) and (c) zoom into the third-order (TE3) mode. The first-order TE (TE1) mode is also visible at \SI{11.7}{GHz}.


A modified Butterworth–Van Dyke circuit fitting is applied here, with routing parasitics (\(L_s\), \(R_s\)), static capacitance \(C_0\), and motional branches for TE1, TE2, and TE3 modes \cite{10718726}. The fitting yields the extracted circuit parameters summarized in Table~\ref{tab:fit_params}. Both the amplitude and phase of the admittance are used for modeling to achieve higher accuracy. The fit is weighted to prioritize the \SI{60}{GHz} resonance and shows reduced agreement below \SI{40}{GHz}, likely due to additional frequency-dependent parasitics in the two-port pad/interconnect structure beyond the present compact model, which will be addressed in future work through improved test structures and modeling. A coupling of $k^{2}=\SI{0.8}{\%}$ is achieved. Here, a rather large fitted $Q$ value of over 100 is obtained for the third-order TE mode, but given the large $R_s$, this value is less reliable. Instead, the phase-derived Bode quality factor\cite{9593620} peaks at $Q=50.91$ at \SI{63.75}{GHz} [Fig.~\ref{fig:admittance_meas}(d)], leading to an overall high $f \cdot Q$ product of $3.13 \times 10^{12}$.

\begin{table}[t]
\caption{Extracted equivalent-circuit parameters.}
\label{tab:fit_params}
\centering
\setlength{\tabcolsep}{2.8pt}%
\renewcommand{\arraystretch}{1.03}%
\begin{ruledtabular}
\begin{tabular}{c c c c c c}
Sym. & Param. & Value & Sym. & Param. & Value \\
\hline
$f_{s1}$  & TE1 res.   & \SI{11.72}{GHz} & $f_{s3}$ & TE3 res.   & \SI{62.59}{GHz} \\
$k_1^{2}$ & TE1 coupl. & \SI{5.57}{\percent} & $k_3^{2}$ & TE3 coupl. & \SI{0.8}{\percent} \\
$Q_1$     & TE1 fit Q  & 6             & $Q_3$     & TE3 fit Q  & 125 \\
$f_{s2}$  & TE2 res.   & \SI{40.38}{GHz}  & $C_0$     & Static Cap.       & \SI{45}{fF} \\
$k_2^{2}$ & TE2 coupl. & \SI{3.34}{\percent} & $L_s$     & Series Ind.       & \SI{0.06}{nH} \\
$Q_2$     & TE2 fit Q  & 15             & $R_s$     & Series Resist.    & \SI{52}{\ohm} \\
\end{tabular}
\end{ruledtabular}
\end{table}

\begin{figure}[t]
    \centering
    \includegraphics[width=0.9\columnwidth]{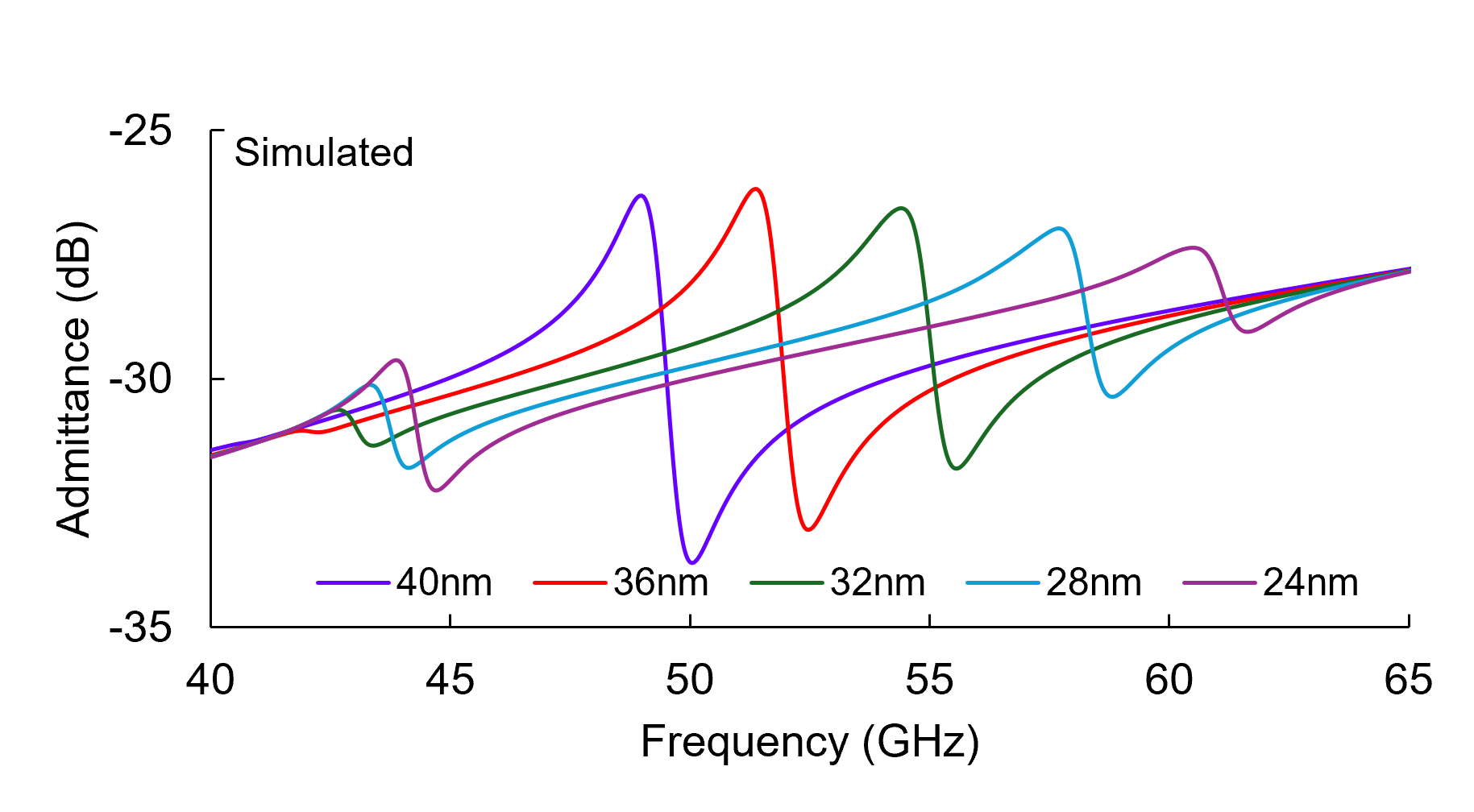}
    \caption{Post-fabrication FEA matching used to estimate top \ce{Pt} thickness. Simulated $f_{s2}$, and $f_{s3}$ are largely overlaid with measurement.}
    \label{fig:post_fea}
\end{figure}

Compared with simulation, the series resonance is loaded by the routing series resistance ($R_{s} = 52~\Omega$) and series inductance ($L_{s} = 0.06~\mathrm{nH}$), which are crucial for fitting the high-frequency response \cite{10.1063/5.0275691, zhang2024acoustic, kramer202257}, This is a common issue for mmWave BAW, as the electrodes are thin.\cite{cho2025challengesscalingscalnbulk} At \SI{62.6}{GHz}, the practical skin depth in thin-film Al is comparable to a few hundred nanometers, so increasing the routing thickness beyond \SI{300}{nm} should still reduce interconnect loss; further reductions in extracted series resistance are expected from lowering thin-film resistivity and improving the Pt–routing metal interface (e.g., in-situ cleaning and alternative routing metals such as Au), which will be explored in future work. Potential methods to address the challenges have been reported, including metal Bragg reflectors \cite{11083120} and multiple resonators in series for reduced impedance \cite{10307494}. Here, as a prototyping effort toward the highest-frequency SMR, we have not included the design, but we will consider these in future iterations. 

Additionally, a second-order TE (TE2) mode appears at \SI{40.4}{GHz}, indicating a thickness mismatch between the top and bottom \ce{Pt} electrodes that imperfectly cancels even-order modes. The measured $k^{2}$ for the third-order TE is also lower than that in the FEA, implying a similar fabrication-related deviation. By matching the measured $f_{s1}$, $f_{s2}$, and $f_{s3}$ with post-fabrication simulations (Fig.~\ref{fig:post_fea}), we estimate a top \ce{Pt} thickness of \SI{28}{nm}, where the third-order mode shifts toward 60 GHz, while the second-order mode starts emerging around 43 GHz due to structure asymmetry. Although the realized operating point is offset from the designed \SI{50}{GHz} target, the $\sim\!60\%$ stopband of the \ce{SiO2}/\ce{Ta2O5} mirror maintains robust confinement despite this offset. Variations in effective longitudinal phase velocity and boundary conditions in thin films, arising from composition- and strain-dependent elastic constants in \ce{ScAlN} and deposition-dependent density/modulus in the reflector, could also account for the shift. Nevertheless, the dominant tone at \SI{62.6}{GHz} remains, to our knowledge, the highest SMR frequency reported to date (Table~\ref{tab:soa}). Future work will back-extract thin-film constants and co-optimize electrode thickness and reflector termination phase to achieve a resonance with tighter tolerance in the desired band \cite{11083120}.

Finally, Table~\ref{tab:soa} benchmarks this work against recent acoustic resonators. While low- to mid-GHz devices often report larger $k^{2}$ and $Q$, our SMR shifts the operating point into the mmWave regime, achieving the highest frequency in the set and coupling comparable to other $\ge\!50$~GHz demonstrations.  Compared with SMRs employing Al/W metal reflectors at similar frequencies\cite{10307494, 11083120}, the lower $Q$ observed here may result from a combination of residual leakage due to reflector bandgap mis-centering (from thin film property deviations) and intrinsic losses/electrical loading in the Pt/ScAlN/Pt cavity and routing. Moreover, because both longitudinal and shear components could contribute to substrate leakage in SMRs\cite{5610561}, a detailed loss decomposition and reflector material comparison aimed at improving confinement of both waves will be pursued in future work. The $k^{2}$ could be improved by future optimization of the fabrication process and the application of other metals, such as iridium (Ir),\cite{4409733} ruthenium (Ru),\cite{ueda2005high} and titanium (Ti).\cite{11083120} The resulting $f \cdot Q$ product is primarily limited by high-frequency loss. Structures reported recently, such as periodically poled piezoelectric films (P3F), could be leveraged to reduce losses by using larger resonant cavities.\cite{10.1063/5.0275691, 10600386, 10188141, cho202423} Nevertheless, the high-contrast \ce{SiO2}/\ce{Ta2O5} Bragg mirror and buried-electrode thickness-field excitation provide a practical route to mmWave operation.

In conclusion, we have demonstrated a \ce{Sc_{0.3}Al_{0.7}N} solidly mounted resonator operating at \SI{62.6}{GHz}, enabled by a \ce{SiO2}/\ce{Ta2O5} Bragg reflector and a buried \ce{Pt} bottom electrode that supports thickness-field excitation. The fabricated devices exhibit $k^{2}=\SI{0.8}{\%}$ and a Bode $Q$ of 50.91, which is, to our knowledge, the highest operating frequency reported for an SMR to date. These measurements validate the design methodology and establish a practical path toward compact, low-loss mmWave acoustic components. Future work will target higher $Q$ via loss reduction in the piezoelectric film and electrodes, higher $k^{2}$ through stack/electrode co-optimization, and composition tuning of \ce{ScAlN}, alongside integration into filter prototypes for next-generation RF front ends.

\section*{SUPPLEMENTARY MATERIAL}
See the supplementary material for (i) finite-element simulations showing the dependence of Bragg-reflector confinement on the number of Bragg reflector layers, and (ii) the measured and fitted complex admittance response of the reported resonator.

\begin{acknowledgments}
This work was supported by the Defense Advanced Research Projects Agency (DARPA) under the Compact Front-End Filters at the Element-Level (COFFEE) program No.\ HR0011-22-2-0031 and by the National Science Foundation (NSF) under CAREER Award No.\ 2339731.
The authors thank Dr.\ Ben Griffin, Dr.\ Todd Bauer, and Dr.\ Zachary Fishman for insightful discussions.

\end{acknowledgments}

\section*{Author Declarations}
\subsection*{Conflicts of Interest}
The authors have no conflicts to disclose.

\subsection*{Author Contributions}
\noindent\textbf{Yinan Wang:} Conceptualization; Investigation; Writing draft.
\textbf{Byeongjin Kim:} Investigation; Data curation; Writing draft. 
\textbf{Nishanth Ravi:} Materials/stack analysis. 
\textbf{Kapil Saha:} ScAlN deposition.
\textbf{Supratik Dasgupta:} Bragg reflector deposition.
\textbf{Vakhtang Chulukhadze:} Investigation.
\textbf{Eugene Kwon:} Materials/stack analysis.
\textbf{Lezli Matto:} Materials/stack analysis.
\textbf{Pietro Simeoni:} ScAlN deposition.
\textbf{Omar Barrera:} Measurements.
\textbf{Ian Anderson:} Modeling; Writing—review \& editing.
\textbf{Tzu-Hsuan Hsu:} Modeling; Writing—review \& editing.
\textbf{Jue Hou:} Bragg reflector deposition.
\textbf{Matteo Rinaldi:} Supervision.
\textbf{Mark S. Goorsky:} Supervision.
\textbf{Ruochen Lu:} Conceptualization; Methodology; Supervision; Writing \& editing.

\section*{Data Availability Statement}
The data that support the findings of this study are available from the corresponding authors upon reasonable request.

\bibliographystyle{aipnum4-1}
\section*{References}
\bibliography{main}

\end{document}